\documentstyle[prl,aps,multicol,epsf,psfrag,graphicx,psfig]{revtex}

\title {\Large\bf
  Entanglement, Elasticity and Viscous Relaxation of Actin Solutions}

\author{B. Hinner$^*$, M. Tempel$^*$, E.  Sackmann$^*$, K. Kroy$^\#$, E.
  Frey$^{\#,\S}$}

\address{$^*$Institut E22, Biophysik, $^\#$Institut f\"ur Theoretische Physik,
  Technische Universit\"at M\"unchen, James-Franck-Stra\ss e,\\
  85747 Garching, Germany; $^\S$present address: Physics Department, Havard
  University, Cambridge, MA 02138}

\begin{document}

\maketitle

\begin{abstract}
  We have investigated the viscosity and the plateau modulus of actin solutions
  with a magnetically driven rotating disc rheometer. For entangled solutions
  we observed a scaling of the plateau modulus versus concentration with a
  power of 7/5.  The measured terminal relaxation time increases with a power
  3/2 as a function of polymer length. We interpret the entanglement transition
  and the scaling of the plateau modulus in terms of the tube model for
  semiflexible polymers.
\end{abstract}


\begin{multicols}{2}
  Networks of semiflexible macromolecules are major constituents of biological
  tissue. 
  There is experimental evidence
  \cite{smith-finzi-bustamante:92,goetter-etal:96,amblard-etal:96,%
    riveline-etal:97} that certain aspects of biologically important
  macromolecules, such as DNA and actin, are well described by the minimal
  theoretical model of a semiflexible macromolecule, also known as the {\em
    wormlike chain\/} model.  This model represents the polymer as a smooth
  inextensible contour with an energy cost for bending and includes ideal
  flexible chains as a limiting case.  The bending modulus of the single
  molecule can be expected to be constitutive also for the {\em collective\/}
  mechanical properties of gels and sufficiently concentrated solutions of
  semiflexible polymers. (Recently, possible contributions from twist have also
  been discussed \cite{maggs:97}.)  However, very little is known about how
  semiflexible polymers build up statistical networks, and how the macroscopic
  stresses and strains are mediated to the single molecules in such networks.
  This is also known as the {\em entanglement problem}. 
  In this Letter, we report on experiments performed with a magnetically driven
  rotating disc rheometer, which elucidate some important aspects of the
  entanglement problem. The systems under scrutiny are {\em in vitro\/}
  polymerized actin solutions of various concentrations $c$ and average polymer
  lengths $L$. {\em Actin\/} \cite{kaes-etal:96} forms large semiflexible
  polymers with a persistence length $\ell_p$ of about 17 $\mu$m
  \cite{ott-magnasco-simon-libchaber:93,gittes-etal:93} (comparable to typical
  filament lengths in our experiments), and is the most abundant cytoskeletal
  element in most eucariotic cells.
  We have analyzed the transition from the dilute to the semidilute phase (the
  entanglement transition) as a function of polymer length and concentration.
  The data can be interpreted in terms of a virial expansion for effective
  ``tubes''.  For entangled solutions we observed a scaling of the plateau
  modulus $G^0$ versus actin concentration $c$. This is compared with various
  theoretical predictions
  \cite{odijk:83,semenov:86,mackintosh-kaes-janmey:95,isambert-maggs:96,%
    satcher-dewey:96,kroy-frey:96}.  Lastly, we analyzed the dependence of the
  zero shear rate viscosity on polymer length, which exhibits a much weaker
  length dependence than one would expect theoretically from work by Odijk
  \cite{odijk:83} and Doi \cite{doi:85}.
  
\begin{figure}[bt]
  \narrowtext
  \centerline{  \epsfxsize=0.75\columnwidth \epsfbox{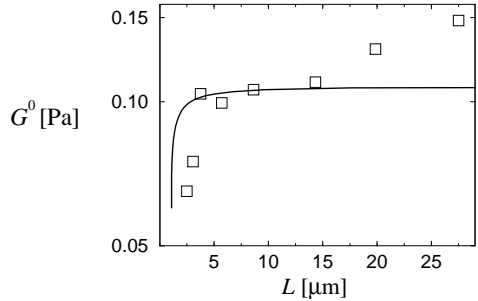}}
     \caption{The plateau modulus above the entanglement transition as a
       function of polymer length for constant monomeric actin concentration
       $c=1.0$ mg/ml. The solid line corresponds to Eq.~(\ref{eq:virial}) with
       $\xi_1=0.38$. The increase of $G^0$ for large $L$ is not yet fully
       understood.}
     \label{fig:g0_L}
\end{figure}

\begin{figure}[bt]
  \narrowtext
  \centerline{ \epsfxsize=0.75\columnwidth  \epsfbox{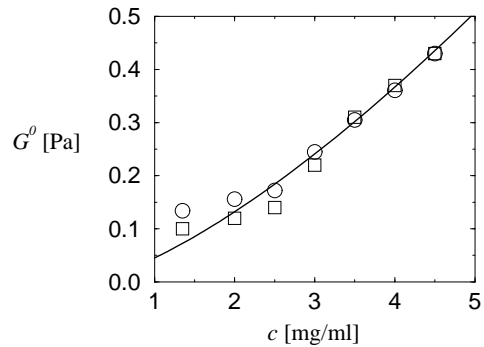}}
  \caption{The plateau modulus near the entanglement transition as a
    function of polymer concentration for short rod-like actin filaments
    ($L=1.5 \;\mu$m). Two different methods were used to extract $G^0$ from the
    data (see main text). Also shown is the theoretical prediction,
    Eqs.~(\protect\ref{eq:virial}) for $\xi_1=0.47$. Theoretically the
    transition is expected at $c^*=0.68$ mg/ml.}
  \label{fig:g0_c2}
\end{figure}

\begin{figure}[bt]
  \narrowtext 
  \centerline{\epsfxsize=0.65\columnwidth \epsfbox{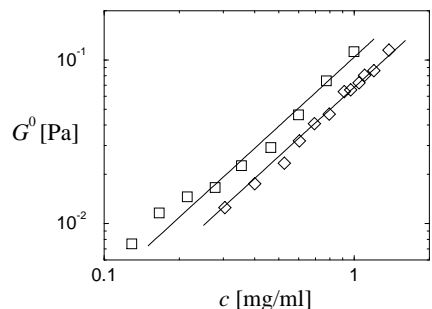}}
     \caption{Concentration dependence of the plateau modulus of pure actin
       ($\Box$) and actin with a small amount of gelsolin ($r_{\rm AG}=6000:1$)
       corresponding to an average actin filament length $L=16$ $\mu$m
       ($\Diamond$). The straight lines indicate the power 7/5 corresponding to
       the scaling limit of Eq.~(\protect\ref{eq:virial}) with $\xi_1=0.40$ and
       $\xi_1=0.46$, respectively.}
     \label{fig:g0_c}
\end{figure}

  Actin was prepared as previously described \cite{tempel-etal:96}, and
  purified in a second step using gel column chromatography (Sephacryl S-300).
  Monomeric actin (called G-actin) was kept in G-buffer, consisting of 2~mM
  Imidazol (pH 7.4), 0.2~mM CaCl$_2$, 0.2~mM DTT, 0.5~mM ATP, and 0.005 vol\%
  NaN$_3$. Polymerization was initiated by adding 1/10 of the sample volume of
  10-fold concentrated F-buffer containing 20~mM Imidazol (pH 7.4), 2~mM
  CaCl$_2$, 1~M KCl, 20~mM MgCl$_2$, 2~mM DTT, and 5~mM ATP.  Gelsolin was
  prepared from bovine plasma serum according to Ref.~\cite{cooper-etal:87},
  and stored dissolved in G-buffer at $4^\circ$C for several weeks. The purity
  of the proteins was checked by SDS-PAGE electrophoresis. After staining with
  coomassie blue \cite{laemmli:70} only one single band was detected.  The mean
  length of actin filaments was adjusted by adding gelsolin to G-actin before
  initiating polymerization.  According to results by Janmey {\it et al.}
  \cite{janmey-etal:86} we computed the average actin length from the molar
  ratio $r_{\rm AG}$ of actin to gelsolin as $L\; [\mu\text m]=r_{\rm AG}/370$.
  All measurements were done at room temperature $(20\pm0.1)^\circ$C.  Both
  oscillatory and creep experiments were performed with a magnetically driven
  rotating disc rheometer, as described previously \cite{tempel-etal:96}.  Care
  was taken to keep the strain below $1\%$ to probe linear response.  For
  oscillatory measurements the phase shift between exciting force and observed
  oscillation and the response amplitude were recorded.  From these two
  parameters the dynamic storage and loss modulus (real and imaginary part of
  the stress amplitude divided by the strain amplitude) were obtained for
  frequencies $\omega/2\pi=10^{-4}$ to $10^1$ Hz.  The creep compliance $J(t)$
  was obtained for times $t=10^{-1}$ to $10^4$ s by applying a sudden step
  force to the sample and recording its strain, which is proportional to
  $J(t)$.  In both cases the apparatus was calibrated with purely viscous
  liquids of known viscosities.  A quantitative measure for the elastic
  character of a material is the phase shift. In the limiting case of a purely
  elastic medium the phase shift is zero, in the opposite case of a purely
  viscous liquid the phase shift is $\pi/2$.  Consequently, the sample behaves
  most rubber-like when the phase shift becomes minimal.  Therefore, in
  oscillatory experiments with actin/gelsolin the value of the storage modulus
  at the frequency corresponding to the minimum phase shift was identified as
  the {\em plateau modulus\/} $G^0$.  For actin samples without gelsolin, where
  no minimum in the phase could be observed within the measured frequency
  range, the storage modulus at a fixed frequency in the plateau regime was
  taken as $G^0$.  This does not affect the functional form of $G^0(c)$ but its
  absolute value. (As a consequence, the vertical shift between the two data
  sets shown in Fig.~\ref{fig:g0_c} has no physical significance.)  For the
  circles in Fig.~\ref{fig:g0_c2}, $G^0$ was determined by the minimum phase
  prescription at the highest concentration only, whereas relative shifts of
  $G^0$ at lower concentrations were determined by rescaling to superimpose the
  moduli.  The {\em zero shear rate viscosity\/} $\eta_0$ was obtained from
  measurements of the creep compliance $J(t)$ according to
  $\eta_0^{-1}:=\lim_{t\to\infty}J(t)/t$ \cite{ninomiya:63}.  From creep
  experiments we extracted the frequency dependent moduli (to obtain $G^0$ for
  Fig.~\ref{fig:g0_L}) as described in Ref.~\cite{tempel_phd:96}. It was
  checked that the results agree well with corresponding oscillatory
  measurements \cite{tempel_phd:96}.
  
  Figs.~\ref{fig:g0_L}, \ref{fig:g0_c2}, and \ref{fig:g0_c} show the plateau
  modulus $G^0$ as a function of filament length $L$ and actin concentration
  $c$, respectively. The data in Fig.~\ref{fig:g0_L} clearly indicate a
  transition with increasing length of polymers.  Similar results also have
  been obtained by Janmey \textit{et. al.\ } \cite{janmey-etal:94}, recently.
  At first sight, one might be tempted to attribute this transition to the
  mutual steric hindrance in a solution of \emph{rods} at the overlap
  concentration.  The observed transition is indeed in a parameter regime,
  where the polymer length $L$ is not much larger (about a factor of 5) than
  the mesh size $\xi_m$, and we originally attempted to interpret the data this
  way. Some more thought suggests, however, that there is no transition
  expected for the plateau modulus of stiff rods; a sudden increase in the
  shear modulus near the overlap concentration would not be in accord with the
  virial expansion for the osmotic pressure of rods \cite{onsager:49}, which
  predicts a smooth dependence on $c$ and $L$ below the nematic transition. One
  can hardly imagine the shear modulus of a semidilute solution of rods to be
  larger than the osmotic compression modulus. The solution could easily escape
  the shear stress by local compression. On the other hand, the actin solutions
  in our experiments were below the critical concentration for the nematic
  transition \cite{suzuki-etal:91,coppin-leavis:92}. We can thus conclude that
  the observed sudden increase (Fig.~\ref{fig:g0_L}), and the enhanced
  concentration dependence (Fig.~\ref{fig:g0_c2},~\ref{fig:g0_c}) of $G^0$
  above a certain threshold are related to the semiflexible nature of actin
  filaments.  Their persistence length of about 17 $\mu$m
  \cite{ott-magnasco-simon-libchaber:93,gittes-etal:93}, albeit much larger
  than the typical mesh size $\xi_m \simeq 0.1-1$ $\mu$m of the studied
  networks, can not be assumed to be infinitely large.  Otherwise the data
  would have to obey the prediction of the classical theory for dilute and
  semidilute rods \cite{doi-edwards:86}
  \begin{equation}
    \label{eq:doi_edwards}
    G^0 = 3\nu k_BT/5\;.
  \end{equation}
  Here $\nu=3/\xi_m^2L$ is the polymer number density.  It is not conceivable
  that the sudden steep increase of $G^0$ with polymer length is merely due to
  internal modes of \emph{single} polymers neglected in the theory for stiff
  rods.  Instead we are forced to look for a \emph{cooperative effect}.

  In the
following we attempt to give a simple interpretation of our observations based
on the tube concept for semiflexible polymers developed theoretically by Odijk
\cite{odijk:83} and Semenov \cite{semenov:86} and related to the shear modulus
by Isambert and Maggs \cite{isambert-maggs:96}, recently.  Experimentally, it
has been demonstrated by video-microscopy that in semidilute actin solutions
the filaments are confined to tube-like cages \cite{kaes-strey-sackmann:94}.
These cages severely hinder not only transverse and rotational motions but also
undulations on length scales larger than a certain length $L_e$, called
\textit{deflection length} or \textit{entanglement length}.
Using the wormlike chain free energy one can relate $L_e$ to the tube diameter
$d$ by $L_e^3=\sqrt 2d^2\ell_p$ \cite{odijk:83,kroy:phd}.  On the time scale of
the plateau, modes of wavelength smaller than $L_e$ are already equilibrated.
Hence, on this coarse grained time scale we can think of the polymer solution
in terms of an ensemble of tubes.  If we apply a reasoning similar to that used
for the osmotic pressure of dispersed rods \cite{onsager:49} to the plateau
modulus of \textit{dispersed tubes} of length $L$ and diameter $d$, we can
replace Eq.~(\ref{eq:doi_edwards}) by a virial expansion
\begin{equation}
  \label{eq:virial}
  G^0 = (3/5) \nu k_BT(1+2B_2\nu \dots)\;.
\end{equation}
Here, $B_2$ is the second virial coefficient for the tubes. Higher order terms
are negligible for small volume fraction of the tubes.  (The latter turns out
to be less than 0.1 in our case.) The product $B_2\nu$ counts the average
number of collisions of the tubes, and can thus be used to define a collision
length $L_c:=L/2B_2\nu$ (always two tubes are involved in a collision).
According to Ref.~\cite{onsager:49} the second virial coefficient is given by
the excluded volume $B_2=\pi dL^2/4$ of a tube.  However, to stay consistent
with our assumption that short wavelength modes have already relaxed, we
subtract from $L$ half the collision length at each end to account for the
reduced efficiency of dangling ends in the entanglement process.  The above
relation between the second virial coefficient and the collision length thus
becomes
\begin{equation}
  \label{eq:modified_onsager}
  L/L_c-1 = \pi\nu d(L-L_c)^2\!/2 \;.
\end{equation}
We can determine the still unknown tube diameter $d$ from the following
consistency requirement. Following Onsager's argument for the second virial
coefficient we have to pay a price in free energy of the order of $k_BT$ per
length $L_c$ to add a new tube to the solution. On the other hand, to suppress
thermal undulations of wavelengths larger than $L_e$ the tube has to supply a
confinement energy of the order $k_BTL/\!L_e$ to the enclosed polymer.  Now, if
we want the tube to be a pertinent effective representation of the medium
surrounding a test polymer in the entangled polymer solution, these two
energies should be equal.  We do not actually have to introduce a physical tube
into the solution when adding a polymer.  Hence, for consistency we require
$L_c\equiv L_e$, i.e., the number of mutual collisions of the tubes must equal
the number of collisions of the polymers with their tubes.  For entangled
solutions we thus find $G^0= 9 k_BT/5\xi_m^2L_e$, where far from the
entanglement transition $L_e$ and $d$ take their asymptotic values $L_e \approx
0.58\xi_m^{4/5}\ell_p^{1/5}$ and $d\approx 0.37\xi_m^{6/5}\ell_p^{-1/5}$. The
scaling behavior was predicted by Isambert and Maggs \cite{isambert-maggs:96}
from a different reasoning before.  We also note that it is included as a
limiting case in a more detailed analysis concerned with the calculation of the
absolute value of $G^0$ \cite{wilhelm-frey:97b}.  The corresponding scaling
$G^0(c)\propto c^{7/5}$ of the plateau modulus is indicated by the solid lines
in Fig.~\ref{fig:g0_c}. A much stronger concentration dependence -- as
predicted by a purely mechanical model \cite{satcher-dewey:96} or by a model
with thermodynamic buckling \cite{mackintosh-kaes-janmey:95} -- and the scaling
predicted in \cite{kroy-frey:96} are not in accord with our data.  On the other
hand, the agreement of Eq.~(\ref{eq:virial}) with the data seems to hold beyond
the scaling limit of strong entanglement.  To relate the apparent (theoretical)
volume fraction to the nominal experimental actin concentration $c$ we
introduce the symbol $\xi_1$ for the apparent mesh size $\xi_m$ [$\mu$m] of a
solution with $c=1$ mg/ml.  Solving Eq.~(\ref{eq:modified_onsager}) we predict
the entanglement transition to occur at a concentration $c^*\; [\text{mg/ml}]=
7\cdot 2^{1/4}\ell_p^{1/2}\xi_1^2/3\pi(5L/7)^{5/2}$ (weakly bending rod limit
assumed). The critical concentration is thus theoretically by a factor of
$c^*/\bar c \approx 2.0(\ell_p/L)^{1/2}$ larger than the overlap concentration
$\bar c$.  For the persistence length we assumed $\ell_p=17$ $\mu$m
\cite{ott-magnasco-simon-libchaber:93,gittes-etal:93} for all our fits. The
only free parameter of the theoretical curves in
Figs.~\ref{fig:g0_L},~\ref{fig:g0_c2}, and \ref{fig:g0_c} is thus $\xi_1$. It
was chosen as $\xi_1=0.38$ for Fig.~\ref{fig:g0_L}, $\xi_1=0.47$ for
Fig.~\ref{fig:g0_c2}, and $\xi_1=0.40/0.46$ for the upper/lower line in
Fig.~\ref{fig:g0_c}, in reasonable agreement with the value $\xi_1=0.35$
obtained independently by FRAP (fluorescence recovery after photo bleaching)
\cite{schmidt-etal:89}.  The scatter in the value for $\xi_1$ merely reflects
the poor experimental reproducibility of absolute values of $G^0$ for F-actin
solutions.

\begin{figure}[t]
\narrowtext
 \centerline{\epsfxsize=0.8\columnwidth \epsfbox{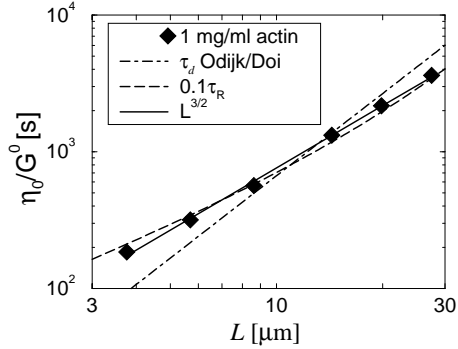}}
     \caption{Terminal relaxation time in the entangled phase 
       for constant actin concentration $c=1.0$ mg/ml. The solid line indicates
       the power $3/2$ and the dashed line is Eq.~(\protect\ref{eq:tau_r}) with
       a numerical prefactor of 0.10.  The dot-dashed line is the
       disentanglement time calculated by Odijk and Doi
       \protect\cite{odijk:83,doi:85}.}
    \label{fig:tau_L}
\end{figure}

Simultaneously with the length dependence of the plateau modulus shown in
Fig.~\ref{fig:g0_L}, we have measured the length dependence of the zero shear
rate viscosity $\eta_0$. The latter is partly due to static effects, namely the
length dependence of the plateau modulus discussed above, and also to dynamics.
The {\em terminal relaxation time} $\tau_r$, the characteristic time scale at
which a polymer solution begins to flow, can be obtained up to a numerical
coefficient from the viscosity via $\tau_r\simeq\eta_0/G^0$.
Fig.~\ref{fig:tau_L} presents such data on the length dependence of $\tau_r$.
Data (not shown) obtained directly from the frequency dependent viscoelastic
moduli by the condition $G'(2\pi/\tau_r)=G^0/2$ or by $\partial
G''(2\pi/\tau_r)/\partial \omega=0$ fall onto the same curve if multiplied by
numerical prefactors $1.0$ and $2.4$ \cite{tempel_phd:96}, respectively. The
mechanism for the terminal relaxation seems obvious from the tube picture
described above.  Viscous relaxation only occurs when the polymers have time to
leave their tube-like cages by Brownian motion along their axis.  The reptation
model that was originally formulated for flexible polymers, was extended to
semiflexible chains by Odijk \cite{odijk:83} and Doi \cite{doi:85}.  These
authors calculated the disengagement time $\tau_d$ for a semiflexible chain
diffusing out of its tube.  However, the data for $\tau_r$ presented in
Fig.~\ref{fig:tau_L} are not in accord with their result for $\tau_d$.  The
dependence of the observed terminal relaxation time $\tau_r$ on polymer length
$L$ is substantially weaker than predicted for $\tau_d$, even in the stiff
limit where $\tau_d=\ell_pL/4D_\| \propto L^2$ (dot-dashed line in
Fig.~\ref{fig:tau_L}), $D_\| = k_BT/2\pi\eta L$ being the longitudinal
diffusion coefficient of the chain in the free draining approximation.
Instead, the solid line in Fig.~\ref{fig:tau_L} corresponds to the scaling law
$\tau_r\propto L^{3/2}$.

A tentative interpretation of the data can be given in terms of a semiflexible
polymer diffusing along a strictly one dimensional path; i.e., not being
allowed to choose between infinitely many new directions at its ends. The
characteristic decay time for self-correlations of the end-to-end vector
$\langle {\mathbf R}(t){\mathbf R\rangle}$ is then given by
\cite{kroy:phd}
\begin{equation}
  \label{eq:tau_r}
  \tau_R = L^4\ell_p^2/D_\| \langle R^2\rangle^2 \approx 
  (L+2\ell_p)^2/4 D_\|\;.
\end{equation}
This presents an upper bound for the terminal relaxation time within the tube
model. As seen from the dashed line in Fig.~\ref{fig:tau_L}, $\tau_R$ (for
$\ell_p= 17$ $\mu$m) is in fact by a factor of ten too large compared to the
data but describes fairly well the length dependence of $\tau_r$. The
restriction to one path implies a very slow decay of conformational
correlations \cite{kroy:phd}. An unusually slow decay of stress (the frequency
dependence of $G'(\omega)$ is still less than linear in the measured frequency
range) is indeed observed, but this might also in part be due to the broad
length distribution of actin \cite{janmey-etal:86}. Clearly, further
investigations are necessary to come to a better understanding of the terminal
regime.

In summary, we were able to measure some important physical properties of
semiflexible polymer solutions with a rotating disc rheometer. We investigated
the plateau modulus and the zero shear rate viscosity of semidilute actin
solutions.  At a certain concentration $c^*$ larger than the overlap
concentration $\bar c$ an increase in the plateau modulus was observed.
We interpreted this entanglement transition as well as the concentration
dependence of the plateau modulus in terms of a tube model that takes into
account the semiflexible nature of the molecules.  For strongly entangled
solutions our data can be characterized by the scaling law $G^0\propto
c^{7/5}$.  We also found a power law dependence of the terminal relaxation time
on polymer length $\tau_r\propto L^{3/2}$, which is substantially weaker than
predicted for the disengagement time by Odijk and Doi.

This work was supported by the Deutsche Forschungsgemeinschaft under Contract
No\@.\ SFB 266. E. F. acknowledges support by a Heisenberg fellowship (FR
850/3-1).  We thank J. Wilhelm for helpful discussions and suggestions, M.
B\"armann for critical comments, B.  Wagner (Fraunhofer-Institut f\"ur
Siliziumtechnologie, Berlin) for the micro-magnets of the rheometer, and our
biochemistry laboratory for preparing the proteins.

\end{multicols}

\end{document}